\pdfoutput=1
\documentclass[twocolumn]{aastex63}

\usepackage{comment}
\usepackage{cancel}
\usepackage{soul}
\received{1 March, 2021}
\revised{June 15, 2021}
\accepted{July 12, 2021}
\submitjournal{AJ}

\shorttitle{Temperature structures in disks}
\shortauthors{Muley et al.}
\begin{document}

\title{Observational Signatures of Planets in Protoplanetary Disks: Temperature structures in spiral arms}

\correspondingauthor{Dhruv Muley}
\email{dmuley@berkeley.edu}

\author{Dhruv Muley}
\affiliation{Department of Physics and Astronomy, University of Victoria, 3800 Finnerty Rd., Victoria, BC V8P 5C2, Canada}

\author{Ruobing Dong}
\affiliation{Department of Physics and Astronomy, University of Victoria, 3800 Finnerty Rd., Victoria, BC V8P 5C2, Canada}
\email{rbdong@uvic.ca}

\author{Jeffrey Fung}
\affiliation{Clemson University, 118 Kinard Laboratory,
Clemson, SC 29634, USA}
\affiliation{Institute for Advanced Study, 1 Einstein Dr., Princeton, NJ 08540, USA}
\email{fung@clemson.edu}
\nocollaboration{3}



\begin{abstract}
High-resolution imaging of protoplanetary disks has unveiled a rich diversity of spiral structure, some of which may arise from disk-planet interaction.
Using 3D hydrodynamics with $\beta$-cooling to a vertically stratified background, as well as radiative-transfer modeling, we investigate the temperature rise in planet-driven spirals. In rapidly cooling disks, the temperature rise is dominated by a contribution from stellar irradiation, 0.3-3\% inside the planet radius but always $<0.5\%$ outside. When cooling time equals or exceeds dynamical time, however, this is overwhelmed by hydrodynamic $P dV$ work, which introduces a ${\sim}10-20\%$ perturbation within a factor of $\sim2$ from the planet's orbital radius. We devise an empirical fit of the spiral amplitude $\Delta(T) = (M_p/M_{\rm th})^c(f_{P dV} e^{-t_{\rm arm}/t_c} + f_{\rm rad})$ to take into account both effects. Where cooling is slow, we find also that temperature perturbations from buoyancy spirals --- a strictly 3D, non-isothermal phenomenon --- become nearly as strong as those from Lindblad spirals, which are amenable to 2D and isothermal studies. Our findings may help explain observed thermal features in disks like TW Hydrae and CQ Tauri, and underscore that 3D effects have a qualitatively important effect on disk structure.

\end{abstract}



\section{Introduction} 
\label{sec:intro}
High-resolution, near-infrared (NIR) scattered-light imaging instruments such as VLT/SPHERE and Gemini/GPI have revealed spiral features in a number of disks, including MWC 758 \citep{Grady2013,Benisty2015} and SAO 206462 \citep{Muto2012,Garufi2013,Stolker2016},
among several others. More recently, spiral structures have been found in ALMA $^{12}$CO observations of TW Hydrae \citep{Teague2019} and CQ Tauri \citep{Wolfer2020}. Various explanations have been proposed for these features \citep{Dong2018}, including e.g. gravitational instability \citep{Toomre1964,Hall2019}, while their ubiquity suggests that at least some of them may have a planet-driven origin. So far, however, direct imaging of planets in protoplanetary disks has proven challenging \citep[e.g.,][]{Keppler2018,Muller2018}, and connections between observed spirals and planets have been circumstantial and inconclusive \citep[e.g.,][]{Ren2018,Wagner2019,Xie2020}.

How, exactly, might planets generate spiral structure in their natal disks? One mechanism, extensively investigated in decades of analytical \citep[e.g.,][]{GT1978,GT1979,Lubow1998,Ogilvie2002} and numerical work \citep[e.g.,][]{Kley1999,Dong2011,Zhu2015,Hord2017,Dong2017,ZhangZhu2020, Ziampras2020}, is wave excitation at Lindblad resonances, where the local orbital frequency is a multiple of the Doppler-shifted forcing frequency of the planet. \cite{GoodmanRafikov2001} studied wake structure and angular-momentum transport in the low-amplitude, linear spirals excited by low-mass planets. \cite{Rafikov2016} and \cite{Arzamasskiy2018} generalized this picture to the arms created by super-thermal mass companions. The superposition of the waves launched at Lindblad resonances constructively interfere to produce multiple arms, as studied by \citep{Bae2018a,Bae2018b,Miranda_2019}.

Less well-studied are the arms launched at buoyancy resonances \citep[e.g.,][]{Zhu2012,Lubow2014,McNally2020,Bae2021}, where the Doppler-shifted planetary frequency commensurates with the local Brunt-Väisälä frequency. This is inherently a 3D effect requiring the solution of an energy equation, and is thus unaccounted for in commonly-used 2D \citep{ZhangZhu2020, Ziampras2020} or 3D isothermal \citep{Dong2017} simulations. \cite{Zhu2012} and \cite{Zhu2015}'s 3D adiabatic simulations found weak buoyancy arms, but given that their temperature structure was vertically unstratified \citep[unlike e.g.,][]{Juhasz2018}, this finding is not necessarily applicable to real disks.

Owing to their isothermality, past hydrodynamical simulations of spiral arms have necessarily focused on \textit{density} perturbations. But while visible in scattered light \citep{Dong2015}, density structures are far less pronounced in gas tracers like $^{12}$CO, which are optically thick for typical disk surface densities. In the present work, we therefore concentrate on the corresponding \textit{temperature} perturbations, which may be more readily apparent in ALMA observations of gas emission.

Gas temperature structure in disks is shaped by two mechanisms. The first is hydrodynamics. As gas passes through spiral arms, compression and expansion subject it to $P dV$ work; this effect are suppressed when the gas-cooling timescale is much shorter than the arm-crossing time, but saturates in the adiabatic limit. \cite{Rafikov2016} finds that for spiral density perturbations up to order-unity, the $P dV$ work done on the gas is mostly reversible; any residual accumulation of heat over successive orbits is due to nonlinear/shock heating \citep{Lyra2016} and is reflected in the gap-opening process, which is not the focus of our work. 

Stellar irradiation also has an important role to play. The atmospheres of passively heated disks
experience direct stellar illumination and thus reach high temperatures. Midplanes
by contrast, are heated only indirectly and are therefore cooler. This gives rise to a vertically stratified equilibrium temperature structure. If a column of disk (in e.g. a spiral) at a given radial-azimuthal location has a greater scale height than its surroundings, it intercepts more starlight and acquires a greater temperature in vertical average, and vice versa.

In what follows, we use 3D adiabatic hydrodynamics with cooling, as well as radiative-transfer simulations, to better understand the temperature structures in spiral arms driven by planets in disks with a realistic vertically stratified temperature structure.

\section{Methods} \label{sec:methods}
We conduct 3D simulations of disk-planet interaction using the GPU-accelerated, Lagrangian-remap hydrodynamics code \texttt{PEnGUIn} \citep{Fung2015Thesis}, which uses the third-order piecewise-parabolic method (PPM) \citep{ColellaWoodward1984} to reconstruct quantities in solving the Riemann problem at cell boundaries. \texttt{PEnGUIn} solves the viscous, compressible Navier-Stokes equations:
\begin{equation}
    \frac{D\rho}{Dt} = -\rho(\nabla \cdot \mathbf{v})
\end{equation}
\begin{equation}\label{eq:momentum_eq}
    \frac{D\mathbf{v}}{Dt} = -\frac{\nabla p}{\rho} + \frac{1}{\rho}\nabla \cdot \mathbb{T} - \nabla \Phi
\end{equation}
\begin{equation}
    \frac{De}{Dt} = -\frac{p}{\rho}\nabla \cdot \mathbf{v} - \Lambda
\end{equation}
where $\rho$ is the mass density, $\mathbf{v}$ the velocity field, $p$ the pressure, $\mathbb{T}$ the viscous stress tensor, $\Phi$ the gravitational potential, $e \equiv p/(\gamma - 1)\rho$ the internal energy per unit mass, and $\Lambda$ the cooling rate. 

We adopt an adiabatic equation of state, $p \equiv c_s^2 \rho/\gamma
%= p_0 (\rho/\rho_0)^\gamma
$, where $c_s$ is the adiabatic sound speed and $\gamma \equiv \partial (\ln p)/\partial (\ln \rho) = 1.4$ is the adiabatic index. $\mathbb{T}$ is proportional to the kinematic viscosity $\nu = \alpha c_s^2 \Omega^{-1}$, where $\Omega$ is the local orbital frequency and $\alpha$ the \cite{ShakuraSunyaev1973} parameter. We choose an $\alpha = 10^{-3}$ in order to prevent the growth of large-scale disk instabilities, such as the Rossby wave instability. This relatively low level of viscosity is also motivated by observations of disks that have generally revealed a low level of turbulence \citep[e.g.,][]{Flaherty2015, Teague2018}.

Our beta-cooling prescription $\Lambda$ relaxes the local sound speed to that of a fixed, vertically-stratified, location-dependent background $c_{s, 0}^2(\mathbf{x})$ 
on a characteristic dimensionless timescale of $t_c \equiv t_{\rm cool, physical} \Omega$:
\begin{equation}
    \Lambda = \frac{t_c^{-1} \Omega}{\gamma - 1}\left(\frac{p}{\rho} - \frac{c_{s, 0}^2(\mathbf{x})}{\gamma} \right)
\end{equation}

The gravitational potential $\Phi$ is given by \cite{Dong2017}
\begin{equation}
    \Phi=-\frac{G M}{1+q}\left[\frac{1}{r}+\frac{q}{\sqrt{r^{2}+R_{\mathrm{p}}^{2}-2 R R_{\mathrm{p}} \cos \phi^{\prime}+r_{\mathrm{s}}^{2}}}-\frac{q R \cos \phi^{\prime}}{R_{\mathrm{p}}^{2}}\right]
\end{equation}
where $G$ is the gravitational constant, $q$ the planet-star mass ratio, $r$ the distance from the origin, $R \equiv r \sin \theta$ the cylindrical radius, $R_{\rm p}$ the radius of the planet, and $\phi' \equiv \phi - \phi_{\rm p}$ the azimuthal separation from the planetary location. Because our simulations are in 3D, our smoothing length $r_s = 0.01875 R_{\rm p}$ is chosen solely to avoid singularity. The planet is fixed on a circular and coplanar orbit with $R_{\rm p} = 1$ in code units. A moderate inclination that keeps the planet within the disk ($i \lesssim h$, where $h$ is the disk aspect ratio) is unlikely to enhance spiral temperature structures as strongly as it does scattered light \citep[e.g.,][]{Kloster2019}, although more substantial inclination would diminish the planet's ability to transmit angular momentum to the disk and consequently reduce spiral visibility. We leave a detailed study on the effect of inclination to future work.

\begin{figure}[h]
    \centering
    \includegraphics[width=0.50\textwidth]{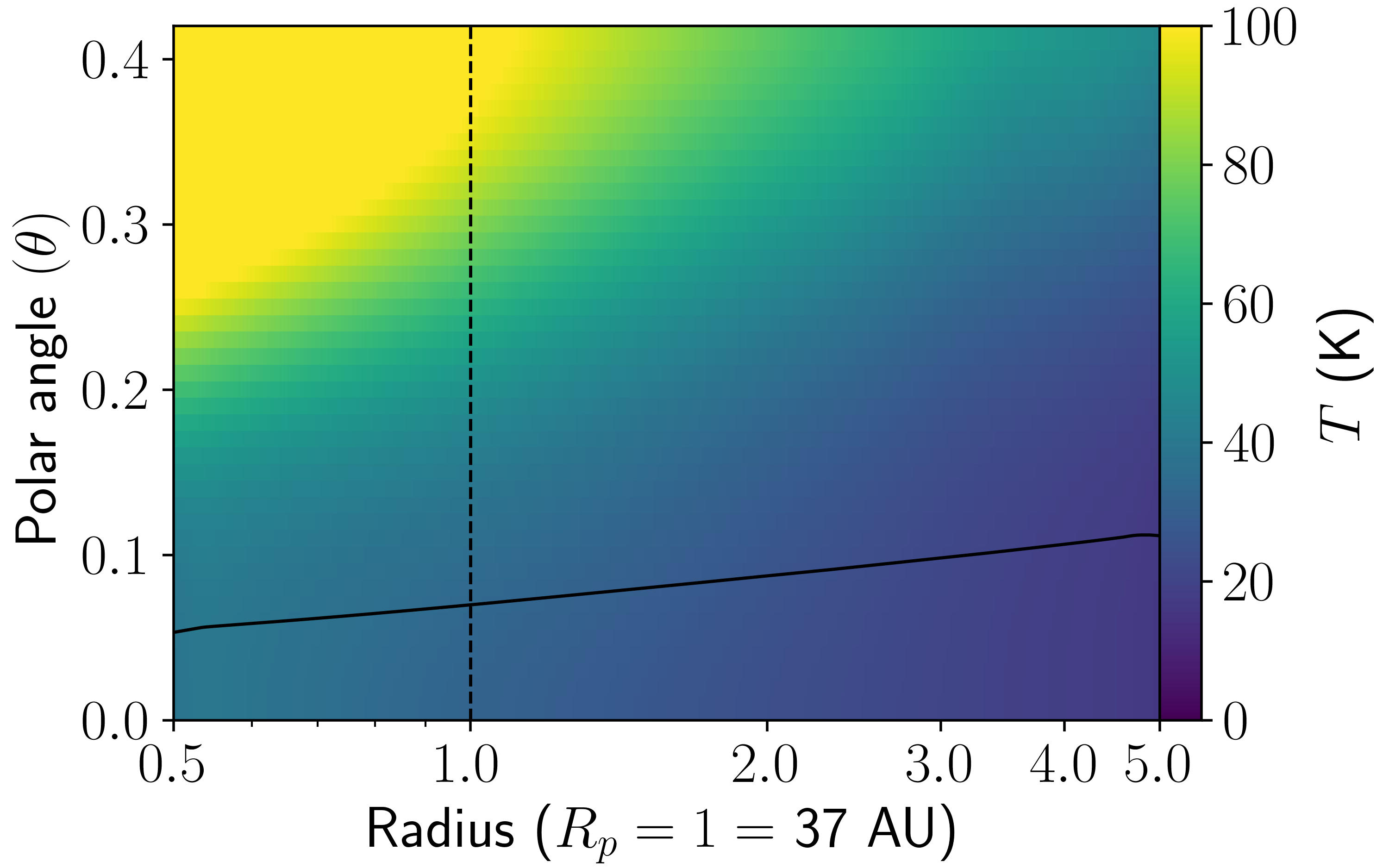}
    \caption{The stratified background temperature of our disk (no planet) obtained from a radiative transfer calculation assuming hydrostatic equilibrium, clearly rising with increasing distance above the midplane. The aspect ratio at a given cylindrical radius, $h(R)$, is plotted as a solid black line. We indicate the location of the planet $R_p = 1 = 37$ AU with a vertical dotted line. 
    }
    \label{fig:temp_profile}
\end{figure}
\subsection{Background temperature profile} \label{sec:diskmodel}

We obtain an axisymmetric background temperature profile, stratified in both radial and vertical directions, using the radiative-transfer code \texttt{HOCHUNK3D} \citep{Whitney2013}. Because radiative transfer is not scale-free, we set fiducial parameters loosely inspired by TW Hya; the planetary location at $R_p = 1$ is scaled to 37 au (where one of the gaps in the system is located, see \cite{Tsukagoshi2016}), and the disk mass set to $30 M_J$ with a dust-to-gas ratio of 0.01 \citep[all interstellar medium grains from][, well-coupled to gas]{Kim1994}. We set the stellar radius to 2.09 $R_\oplus$ and temperature to 4000 K, typical of T Tauri stars. To ensure that the vertical distribution of disk material reflects the temperature profile, we enable the \texttt{HSEQ} mode of \texttt{HOCHUNK3D} \citep{Whitney2013}, which iterates the vertical disk profile after each RT iteration until convergence is reached.

We fit the resulting \texttt{HOCHUNK3D} temperature as described in Appendix \ref{sec:appendix}, and plot the result in Figure \ref{fig:temp_profile}. From here, we find the sound speed as $c_{s, 0}^2(\mathbf{x}) = \gamma k_B T(\mathbf{x})/\mu m_{\rm H}$, taking the mean molecular weight $\mu = 2.34$ as in the minimum-mass solar nebula. 
At the location of the planet, the effective aspect ratio $h_{\rm eff} \equiv c_{s,0}(\mathbf{x})/(\gamma^{1/2} R\Omega) \approx 0.07$. For a vertically isothermal disk in hydrostatic equilibrium, this would equal the true aspect ratio $h$, but in a stratified disk we must define it explicitly as
\begin{equation}
\label{eq:true_sch}
    h(R) \equiv \left(\frac{1}{R^2\Sigma(R)} \int_{-\infty}^{\infty} z^2 \rho(z) dz\right)^{1/2}
\end{equation}

\subsection{Disk setup and tests} \label{sec:setup}
\begin{figure}
    \centering
    \includegraphics[width=0.50\textwidth]{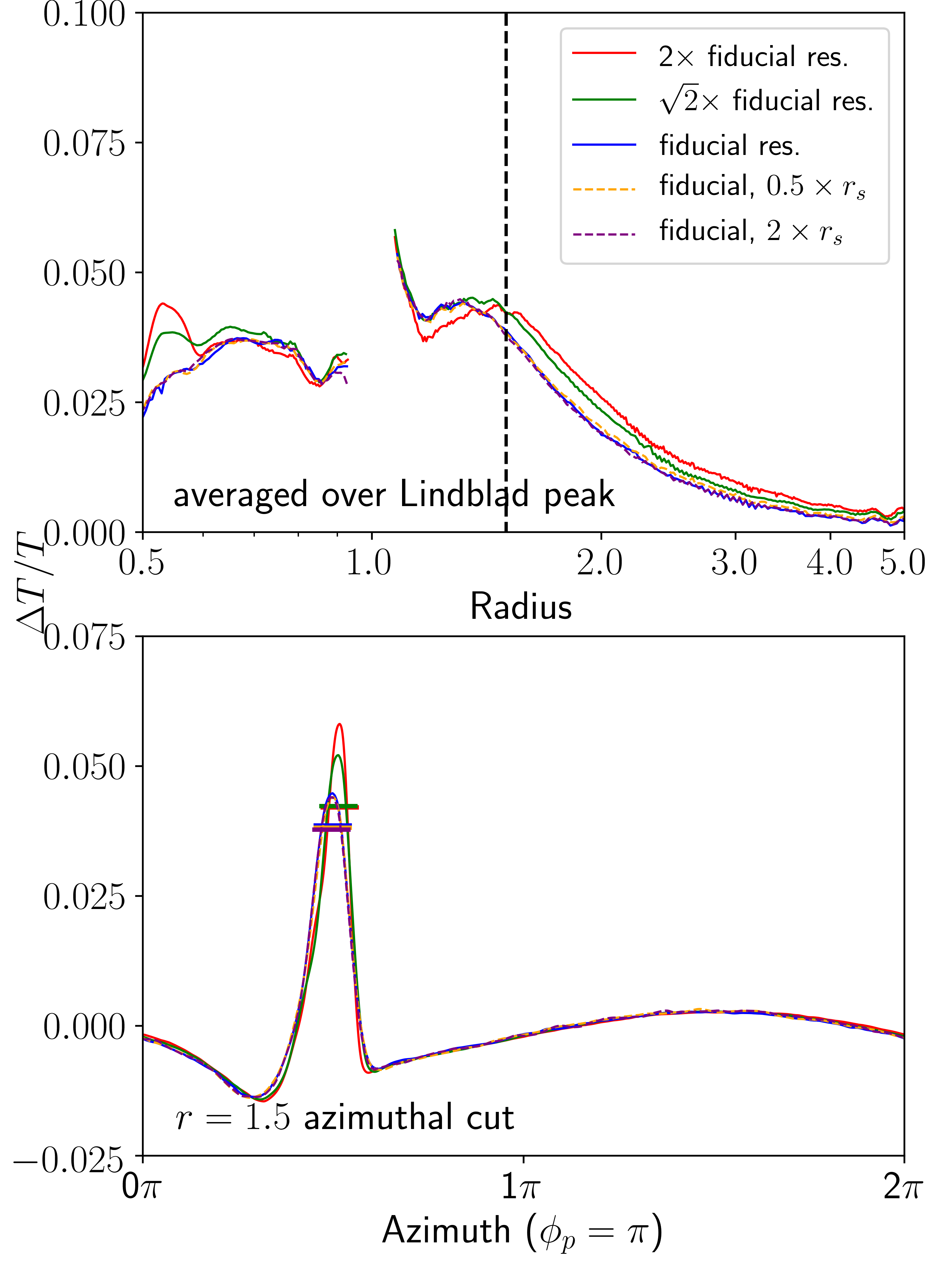}
    \caption{Numerical convergence for the average temperature perturbation along the primary Lindblad spiral (Equation \ref{eq:spiral_def}) as a function of radius \textit{(top)}, and as a function of azimuth in an $r = 1.5$ cut \textit{(bottom)} for
    simulations with $M_p = 50 M_{\oplus}$ and $t_c = 10^2$. In the bottom panel, $(\Delta T/T)_{\rm spiral}$ and the azimuthal range it is averaged over (see Equation \ref{eq:spiral_def}) are indicated by short straight lines. Smoothing length has a strong effect near the planet, but has less impact farther away; resolution convergence is best within a factor 2 of the planet radius, but worsens far away due to numerical diffusion. Much of the disagreement with resolution arises due to wave steepening at $\phi_{\rm peak}$, which has minimal impact on the integrated spiral perturbation.     }
    \label{fig:num_convergence}
\end{figure}

We initialize the disk with a surface density \begin{equation}
    \Sigma = \Sigma_0 \left(\frac{R}{R_p}\right)^{-1}
\end{equation}
Our science simulations use a resolution of $267 (r) \times 683 (\phi) \times 42 (\theta)$, spanning a radial range of $r = \{0.4, 5\}$, a polar range of $\theta = \{\pi/2, \pi/2 - 0.42\}$ (covering $\sim6$ scale heights from the midplane at the planet location)\footnote{Our hydrodynamics assume symmetry about the midplane; in \texttt{HOCHUNK3D}, where this is not presumed, we simply reflect cells across it.}, 
and the full $2\pi$ in azimuth. Cells are spaced logarithmically in the radial direction, but uniformly in the polar and azimuthal directions. This yields a resolution of roughly 7 cells per effective scale height at the location of the planet. 
We use periodic boundary conditions in the azimuthal direction, and reflective boundary conditions in the polar direction; in the radial direction, we use outflow boundaries because appropriate fixed boundary conditions for density, velocity, etc. for our stratified temperature structure lack an analytic expression.

We verify our implementation of beta-cooling with tests at $t_c = 10^{-4}$ and $10^{-6}$; 
these demonstrate typical fractional azimuthal temperature perturbations of order $\sim 10^{-5}$ and $\sim 10^{-14}$, respectively (not shown). In vertical average, however, we see fluctuations in temperature deviating from the background at percent-level;
this is because, even though temperature in any given grid cell deviates little from its background value, mass is redistributed upward and downward in the disk by the spiral density wave, changing the column-averaged temperature. This effect would not be captured in 2D simulations, so we cover it in more detail in the following section.

In the opposite limit of \textit{long} cooling times, it is not immediately clear whether our simulations would reach any sort of steady-state. To test whether they do, we run $t_c = 10^4$ models for each planet mass, and find that they yield essentially identical temperature and density structures to our $t_c = 10^2$ runs. As an additional test, we run a $M_p = 50 M_\oplus$, $t_c = 10^4$ simulation out to 400 orbits, and find that once spiral morphology and amplitude is established at $t \approx 10$ orbits, it remains in steady-state (up to gap-opening) throughout. 
This is because planet-induced spiral arms are patterns that gas enters (to be compressed and heated) and leaves (to expand and cool) over the course of a single orbit, rather than persistent accumulations of gas.

Throughout this work, we define the relative strength along the spine of a Lindblad spiral at a given cylindrical radius $(\Delta T/T)_{\rm spiral}(R)$ as:
\begin{equation}
\label{eq:spiral_def}
    (\Delta T/T)_{\rm spiral}(R) = \frac{1}{2h_{\rm eff, p}}\int_{\phi_{\rm peak}(R) - 2 h_{\rm eff, p}}^{\phi_{\rm peak}(R) + 2 h_{\rm eff, p}} d\phi (\Delta T/\left<T\right>_{\phi})(R, \phi)
\end{equation}
where $\phi_{\rm peak}(R)$ is the azimuthal location of the spiral density peak and $h_{\rm eff, planet} = 0.07$ is the effective aspect ratio at the location of the planet, and $(\Delta T/\left<T\right>_{\phi})(R, \phi)$ is the mass-weighted, fractional azimuthal perturbation in vertically-averaged temperature at a given 2D location:
\begin{equation}
    \left<T\right>_{\phi}(R) = \frac{\int_0^{2\pi} \int_0^{\infty} d\phi dz \rho T}{\int_0^{2\pi} d\phi \Sigma(R, \phi) }
\end{equation}
\begin{equation}
    (\Delta T/\left<T\right>_{\phi})(R, \phi) = \frac{1}{\Sigma(R, \phi)}\frac{ \int_{0}^{\infty} dz \rho T}{\left<T\right>_{\phi}(R, \phi)} - 1
\end{equation}

We define $(\Delta \Sigma/\Sigma)_{\rm spiral}(R)$ and $\Delta \Sigma/\left<\Sigma\right>_{\phi}$ analogously. This formulation is motivated by observational relevance---beam convolution would smear out a point estimate of spiral amplitude.

As an additional check, we run convergence tests in both grid resolution and smoothing length $r_s$ at a fiducial $M_p = 50 M_\oplus$ and $t_c = 10^2$, and display our results in Figure \ref{fig:num_convergence}. Our upper panel, plotting $(\Delta T/T)_{\rm spiral}(R)$, shows a well-converged Lindblad spiral amplitude for radii within a factor of 2 of the planet location. Farther away, however, spiral strength is supported by acoustic propagation rather than resonant driving, and thus becomes degraded by numerical diffusion.

Our lower panel---an azimuthal cut at $r = 1.5$---underscores that within a factor ${\sim}2$ of the planet position, most of the difference between resolutions arises from nonlinear wave-steepening at the peak amplitude location $\phi_{\rm peak}$. While properly capturing spiral amplitude at this point would require resolutions far exceeding what is feasible in 3D global simulations \citep{Dong2011a,Dong2011b}, this only changes the integrated arm amplitude by ${\sim}1.15\times$. As for changing $r_s$, we find that it affects features in the immediate co-orbital region, but otherwise has negligible impact on our results.

\begin{figure}
    \centering
    \includegraphics[width=0.5\textwidth]{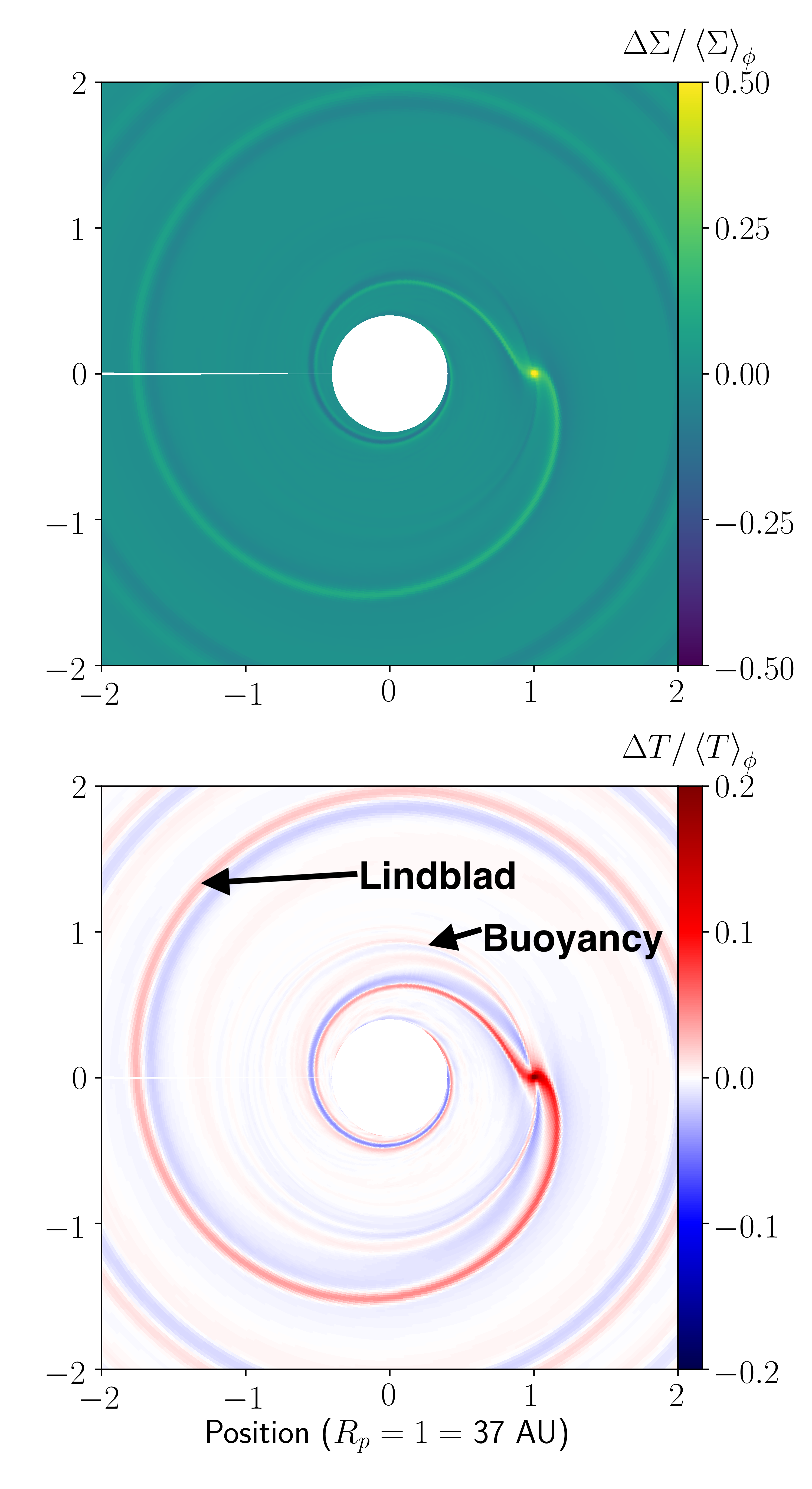}
    \caption{\textit{Top}, a Cartesian view of the vertically-averaged density perturbation for our $M_p = 50 M_\oplus$, $t_c = 10^2$ simulation, at the fiducial cutoff point of $t = 25$ orbits; \textit{bottom}, a vertically-averaged temperature-perturbation map of the same run.
    The color scale for temperature is scaled to $\partial \ln T/\partial \ln \rho=\gamma - 1 = 0.4\times$ that of density, as in the limit of purely adiabatic compression and expansion.}
    \label{fig:cartesian_fiducial}
\end{figure}

Our analysis centers on a grid of 12 models, with cooling timescales $t_c = \{10^{-2}, 10^{0}, 10^2\}$ --- typical of real protoplanetary disks at tens of au \citep[e.g.,][]{Miranda2020,ZhangZhu2020,Ziampras2020} --- and planetary masses $M_p = \{50, 100, 200, 400\} M_\oplus$ (corresponding to planet-star mass ratio $q = \{1.5\times 10^{-4}, 3\times 10^{-4}, 6\times 10^{-4}\, 1.2 \times 10^{-3} \}$, or $q/q_{\rm th} = \{0.48, 0.96, 1.91, 3.83\}$) (where $q_{\rm th} \equiv h_{\rm eff}^3$ is the thermal mass in the disk). At $t = 0$, our simulations are vertically isothermal, with a radially varying temperature given by equation \ref{eq:t_mid} of Appendix \ref{sec:appendix}. We relax to the vertically stratified temperature profile (equation \ref{eq:t_rtheta}) with a cooling timescale of $t_{\rm cool, init} = 0.1 \times 2\pi$ over 10 planetary orbits. At that point, we set the cooling time to its notional value, and initialize the planet, growing it to its final mass over 1 orbit. We run each simulation for 15 more orbits, for a total of 25. We plot Cartesian density and temperature maps for one representative simulation---the $M_p = 50 M_\oplus$, $t_c = 10^2$ run used in our resolution test---in Figure \ref{fig:cartesian_fiducial}.

\begin{figure}
    \centering
    \includegraphics[width=0.5\textwidth]{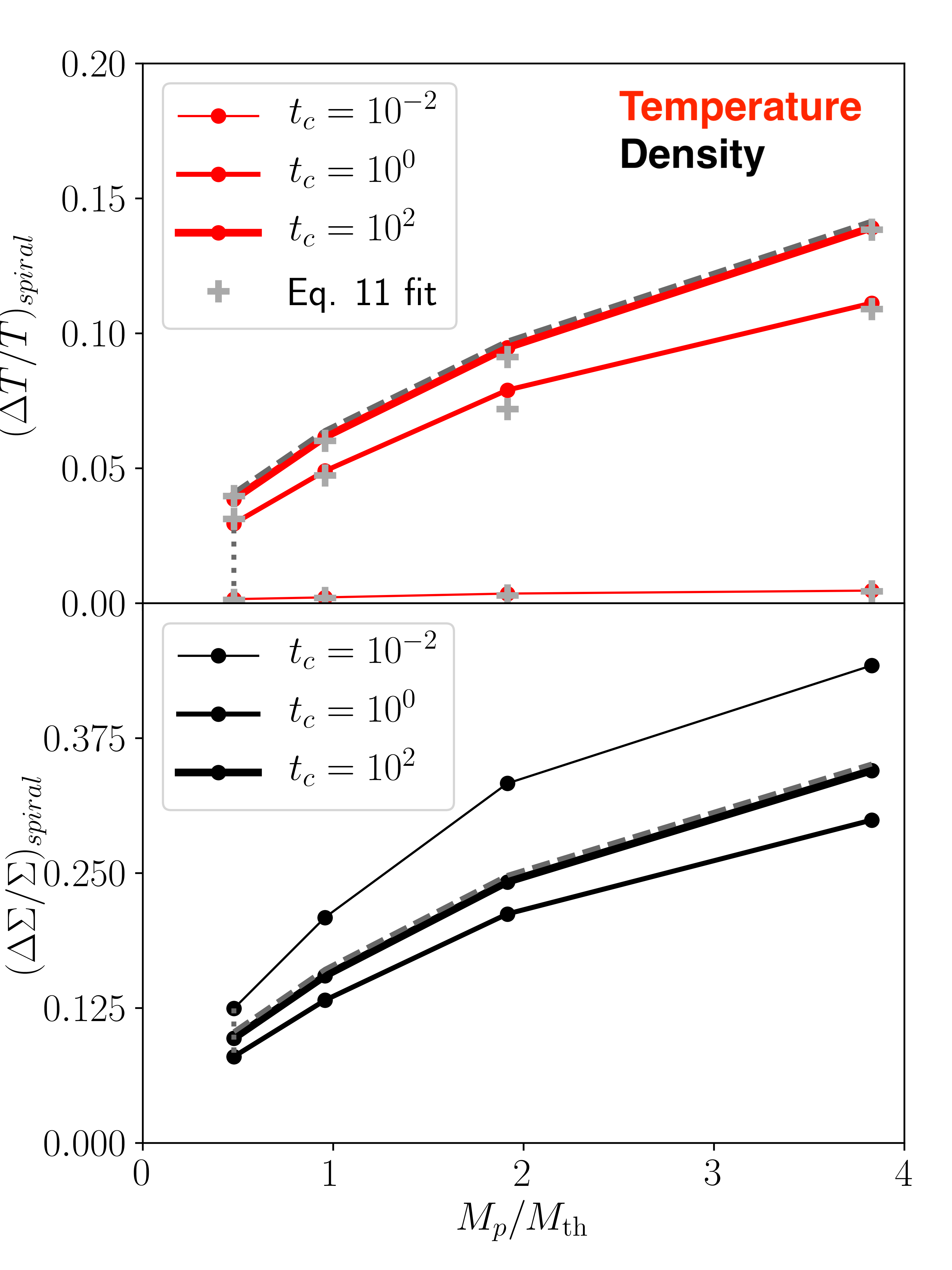}
    \caption{Vertically-averaged Lindblad $(\Delta T/T)_{\rm spiral}$ (Equation~\ref{eq:spiral_def}) and $(\Delta \Sigma/\Sigma)_{\rm spiral}$ amplitudes (analog to Equation~\ref{eq:spiral_def}) for our model grid at $r = 1.5$; we fit the temperature using Equation \ref{eq:empirical_fit}.
    Planet masses are expressed in units of $M_{\rm th}$.
    runs show the strongest density perturbation amplitudes, they show the weakest temperature amplitudes (and are dominated by radiative effects). Temperature amplitude strengthens for our $t_c = 1$ and $10^2$ runs. The dotted and dashed lines, and the points they intersect, correspond to those in Figure \ref{fig:quantitative_p50_perturbation}, reflecting lines of constant cooling time and planet mass.
    }
    \label{fig:aggregate_lindblad_amplitudes}
\end{figure}

\begin{figure}
    \centering
    \includegraphics[width=0.50\textwidth]{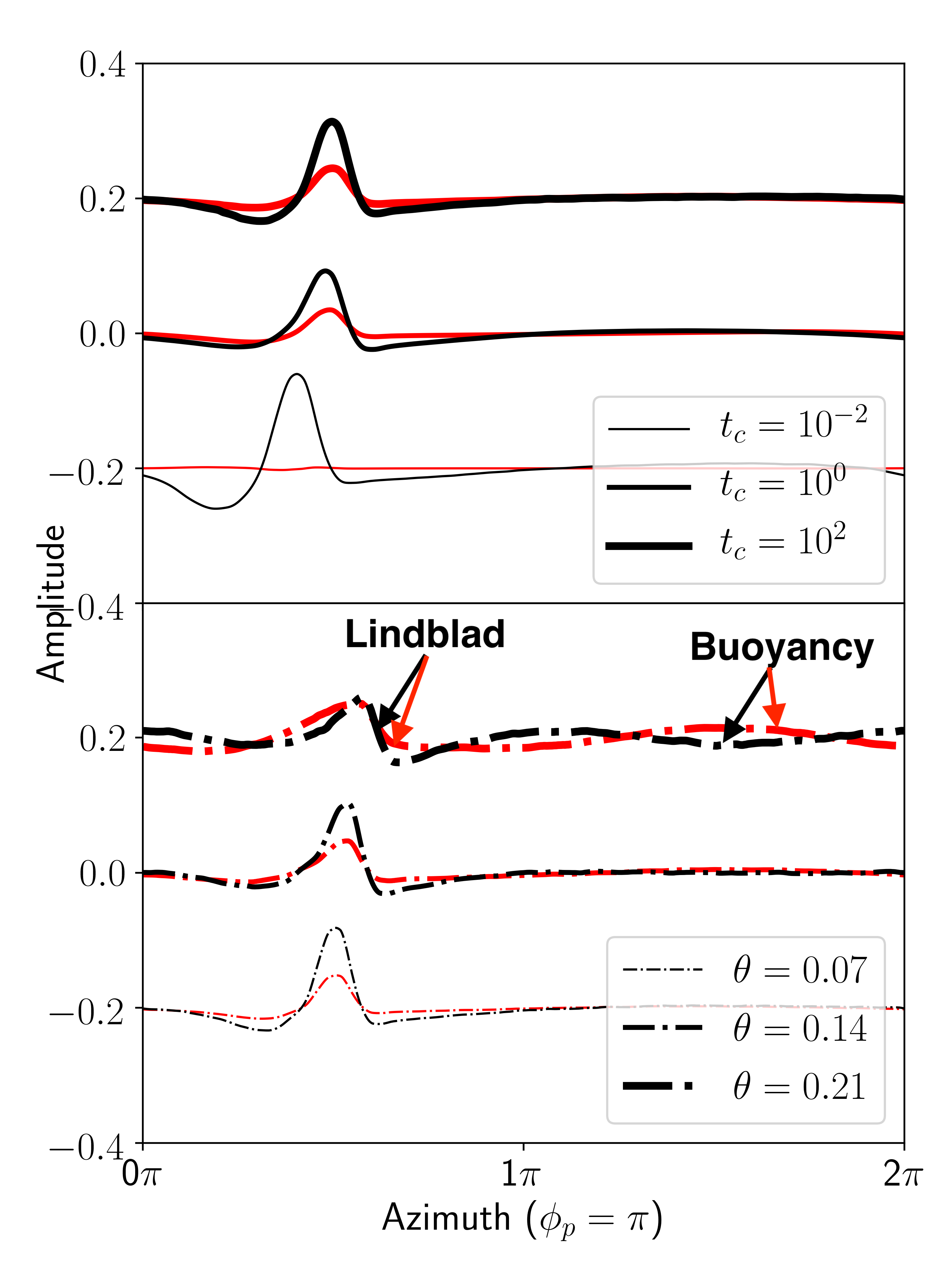}
    \caption{\textit{Above}, vertically-averaged spiral density \textit{(black)} and temperature \textit{(red)} perturbations as a function of azimuth at $r = 1.5$ (see Figures \ref{fig:vert_temp_struct} and \ref{fig:sigma_plot}) for our $M_p = 50 M_\oplus$ models. The density amplitude shrinks by roughly 50\% between $t_c = 10^{-2}$ and $10^2$, while the temperature perturbation increases by an order of magnitude or more. \textit{Below}, perturbations at $r = 1.5$ in our $t_c = 10^2, M_p = 50 M_\oplus$ model, taken at 1, 2, and 3 effective scale heights (at the planet location) above the midplane. Buoyancy spirals grow in amplitude at higher altitude while Lindblad spirals weaken. Different cooling times and scale heights are vertically shifted by 0.2 relative to one another.     }
    \label{fig:quantitative_p50_perturbation}
\end{figure}

\begin{figure*}
    \centering
    \includegraphics[width=1\textwidth]{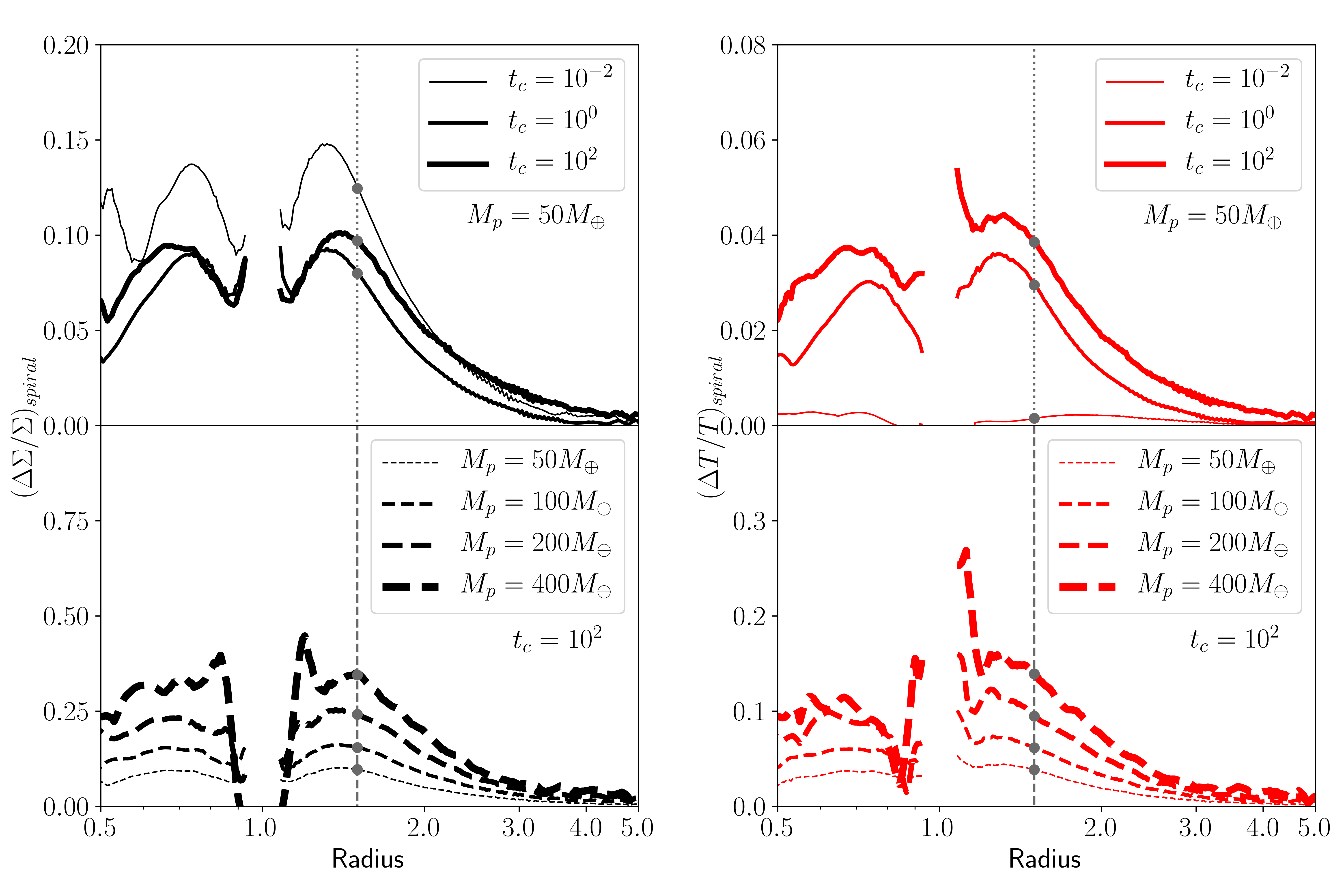}
    \caption{\textit{Left},  $(\Delta \Sigma/\Sigma)_{\rm spiral}$, and \textit{right} $(\Delta T/T)_{\rm spiral}$ (Equation \ref{eq:spiral_def}), as a function of radius. Upper panels fix $M_p = 50 M_\oplus$ and vary $t_c$; lower panels fix $t_c = 10^{2}$ but vary planet mass. }
    \label{fig:quantitative_p50_perturbation2}
\end{figure*}

\section{Results} \label{sec:results}
\subsection{Lindblad spirals} \label{sec:lindblad}
Lindblad spirals are formed by constructive interference of the Fourier modes excited in the disk by the planetary potential. They have a characteristic width of ${\sim}hR$ and a pattern speed of $\Omega(R_p)$, with waves launched at roughly $R_p \pm hR_p$, so one can estimate in the rotating frame that it takes gas parcels at most one dynamical time $t_{\rm dyn} = \Omega^{-1}(R_p)$ to cross through. For the shortest cooling time in our model grid ($t_c = 10^{-2}$), therefore, $P dV$ work by the Lindblad spiral on a gas parcel is dissipated much faster than it is performed, so compression and expansion are nearly isothermal processes. For $t_c \gtrsim 1$, $P dV$ work is retained through an arm crossing, making the compression and expansion effectively adiabatic. We emphasize that, even when the simulation is in ``steady state'', only the spiral pattern is fixed; the underlying gas is not in vertical hydrostatic equilibrium \citep[Fig. 3]{Dong2017}, as gas in the inner spiral arms has significant vertical motion \citep{Zhu2015}. In Figure \ref{fig:aggregate_lindblad_amplitudes}, we plot $(\Delta T/T)_{\rm spiral}$ at $r=1.5$ in our Lindblad spirals for all 12 simulations in our model grid. 

For $t_c = 10^{-2}$, therefore, the temperature in any given grid cell remains essentially unchanged from the prescribed background value; any non-axisymmetry in temperature arises because kinematic effects redistribute material vertically. In the region of the Lindblad arm where a gas parcel is being compressed azimuthally---in anticipation of the spiral density peak---it becomes vertically inflated and thus heats up in vertical average. Following the density peak, gas expands azimuthally and shrinks to the midplane, cooling down in vertical average. This effect is stronger for inner spiral arms ($0.3-3\%$) than for the outer arms ($0.1-0.5\%$) whose amplitudes appear in Figure \ref{fig:quantitative_p50_perturbation}. Results from our test runs at $t_c = 10^{-4}$, not shown here, show quantitatively similar results.

By contrast, when $t_c > 1$, the $P dV$ work done by a Lindblad spiral is no longer dissipated before a gas parcel fully crosses through the arm. As a result, the high-density central spine of the arm becomes hotter adiabatically, while the lower-density regions before and after expand and cool. With rising cooling time, this effect grows to dominate the overall temperature perturbation, causing it to follow the density (rather than its rate of change in azimuth, as in $t_c = 10^{-2}$ runs). As shown in the lower panel of Figure \ref{fig:quantitative_p50_perturbation}, this picture holds at all altitudes in the disk, although increasing sound speed and distance from the planet widen and somewhat weaken the spiral perturbation.

Motivated by these considerations, we present the following fit for the vertically averaged spiral temperature amplitudes as a function of both $M_p$ and $t_c$ plotted in the upper panel of Figure \ref{fig:aggregate_lindblad_amplitudes} for fiducial radius $R/R_p = 1.5$:
\begin{equation}
\label{eq:empirical_fit}
    (\Delta T/T)_{\rm spiral, fit} = \left(\frac{M_p}{M_{\rm th}}\right)^{c}\left[f_{P dV} e^{-t_{\rm arm}/t_{c}} + f_{\rm rad}\right]
\end{equation}
where $c = 0.6$ is the power-law exponent of the temperature curves, $f_{P dV} = 0.06$ is the temperature perturbation arising from $P dV$ work in the adiabatic limit, $f_{\rm rad} = 0.002$ that from vertical material redistribution in the stratified temperature structure, and $t_{\rm arm} = 0.25$ a characteristic timescale for compression and expansion of material in the spiral arms. Differences between fitted and simulated amplitudes are typically ${\lesssim}0.05 \times (\Delta T/T)_{\rm spiral, fit}$; we note that a more extensive parameter survey in the future may help further refine this.

At any given planet mass, the temperature perturbation at any given planet mass is weakest at $t_c = 10^{-2}$ and strongest at $t_c = 10^2$. As visible in our Figure \ref{fig:quantitative_p50_perturbation2}, however, density perturbation is non-monotonic in regions at least several scale heights from the planet---strongest for $t_c = 10^{-2}$, but weakening at $t_c = 1$ before regaining some strength at $t_c = 10^2$ \citep{Ziampras2020,ZhangZhu2020}. The analytical study of \cite{Miranda2020} investigates this in detail, finding (in the linear limit) a radial separation between wavecrests set by 
\begin{equation}
\label{eq:re_k}
    \textup{Re}(k_m) \approx \frac{\left|\Omega^2 - \tilde{\omega}^2 \right|^{1/2}}{c_{\rm s}}
\end{equation}
and an amplitude damping rate
\begin{equation}
    \textup{Im}(k_m) = \frac{(\gamma - 1)\Omega\tilde{\omega}t_c}{2(\Omega^2 + \gamma \tilde{\omega}^2 t_c^2)} \textup{Re}(k_m)
\end{equation}
where $\tilde{\omega} \equiv \pm m(\Omega - \Omega_p)$ is the Doppler-shifted forcing frequency of the planet, times the azimuthal wavenumber $m$. 

Physically, $\textup{Im}(k_m)$ represents the energy lost as beta-cooling erodes the temperature component of the wave over the course of each oscillation. At all Lindblad resonances $\Omega_m = \Omega_p(1 \mp 1/m)$, these losses are maximized for a $t_{c, \rm crit} \approx \gamma^{-1/2} \approx 1$, meaning waves are suppressed at their launching points. Cooling times longer and shorter than this, however, allow Lindblad waves to propagate more freely through the disk.

In Appendix \ref{sec:temp_dens_plots}, we present vertically-averaged galleries of 2D $\Delta T/\left<T\right>_{\phi}$ (Figure \ref{fig:vert_temp_struct}) and $\Delta \Sigma/\left<\Sigma\right>_{\phi}$ maps (Figure \ref{fig:sigma_plot}), as well as $(r, \theta)$ cuts of both temperature and density normalized to the azimuthal average (Figure \ref{fig:temp_dens_aziavg_diff}). These provide a visual and intuitive understanding of the spiral perturbation for all simulations in our model grid.

\subsection{Buoyancy-resonance spirals} \label{sec:buoyancy}
Buoyancy waves become apparent in temperature plots for $t_c \gtrsim 1$ and saturate in strength at $t_c = 10^2$. In each simulation, the vertically-averaged temperature structure reveals at least three sets of buoyancy spirals, corresponding to different azimuthal tidal forcing wavenumber $m$ \citep{Zhu2012,Zhu2015}. Whereas for low planet masses and short cooling times these perturbations are weak (${\sim}1-2.5\%$), they become stronger and substantially distorted in the opposite limit, wrapping around the full $2\pi$ in azimuth and intersecting the Lindblad-resonance arm. In all cases with $t_c \geq 1$, we find buoyancy perturbation strengths several orders of magnitude stronger than in \cite{Zhu2015}, who did not use a stratified background temperature.

The Lindblad spiral is strongest close to the midplane, widening and weakening in the disk atmosphere where sound speed is higher. Buoyancy spirals, however, are weak near the midplane but stronger in the disk atmosphere, the region typically probed by gas tracers such as $^{12}$CO. This is made quantitatively clear in the lower panel of Figure \ref{fig:quantitative_p50_perturbation}. Buoyancy causes hot material to rise and expand while pushing cold material to fall and contract, leaving pressure constant (at fixed altitude); consequently, in the long-$t_c$ limit the ratio between temperature and density perturbations is $\approx-1$ for buoyancy spirals, whereas for Lindblad spirals (an adiabatic phenomenon) the ratio is $\approx (\gamma - 1) = 0.4$.

Previous work \citep[e.g.,][]{Zhu2015,McNally2020,Bae2021} has established that buoyancy resonances occur where the Doppler-shifted forcing frequency of the planet commensurates with the local Brunt-Väisälä frequency:
\begin{equation}
    \pm m(\Omega - \Omega_p) = N
\end{equation}
where in a protoplanetary disk with $g = \Omega^2 z$,
\begin{equation}
    N^2 = \Omega^2\left[\frac{z}{T}\frac{\partial T}{\partial z} - \frac{\gamma - 1}{\gamma}\frac{z}{P}\frac{\partial P}{\partial z}\right]
\end{equation}

Because our disks are vertically stratified, the first (temperature) term is always present, resulting in very weak buoyancy arms close to corotation even with a $t_c = 10^{-2}$. For longer cooling times (equivalently, the effective $\gamma$ rising from 1 to its notional value of 1.4), the second (pressure) term makes resonances diagonal in the upper disk atmosphere, while the first creates an additional bend at the interface between disk midplane and atmosphere, where temperature is changing rapidly. For super-thermal planets, buoyant temperature perturbations are sufficiently strong as to change $N$ and alter the resonance location, creating coupling between different buoyancy modes that manifests as merging and splitting of buoyant arms.

We stress that while the Brunt-Väisälä frequency is well behaved as $\gamma \rightarrow 1$ from above, it is undefined for a truly isothermal equation of state, for which $\gamma = 1$ and the Navier-Stokes energy equation is degenerate. Thus there are no buoyant perturbations in simulations like \cite{Juhasz2018}, even though the background temperature is vertically stratified.

\begin{figure*}
    \centering
    \includegraphics[width=0.98\textwidth]{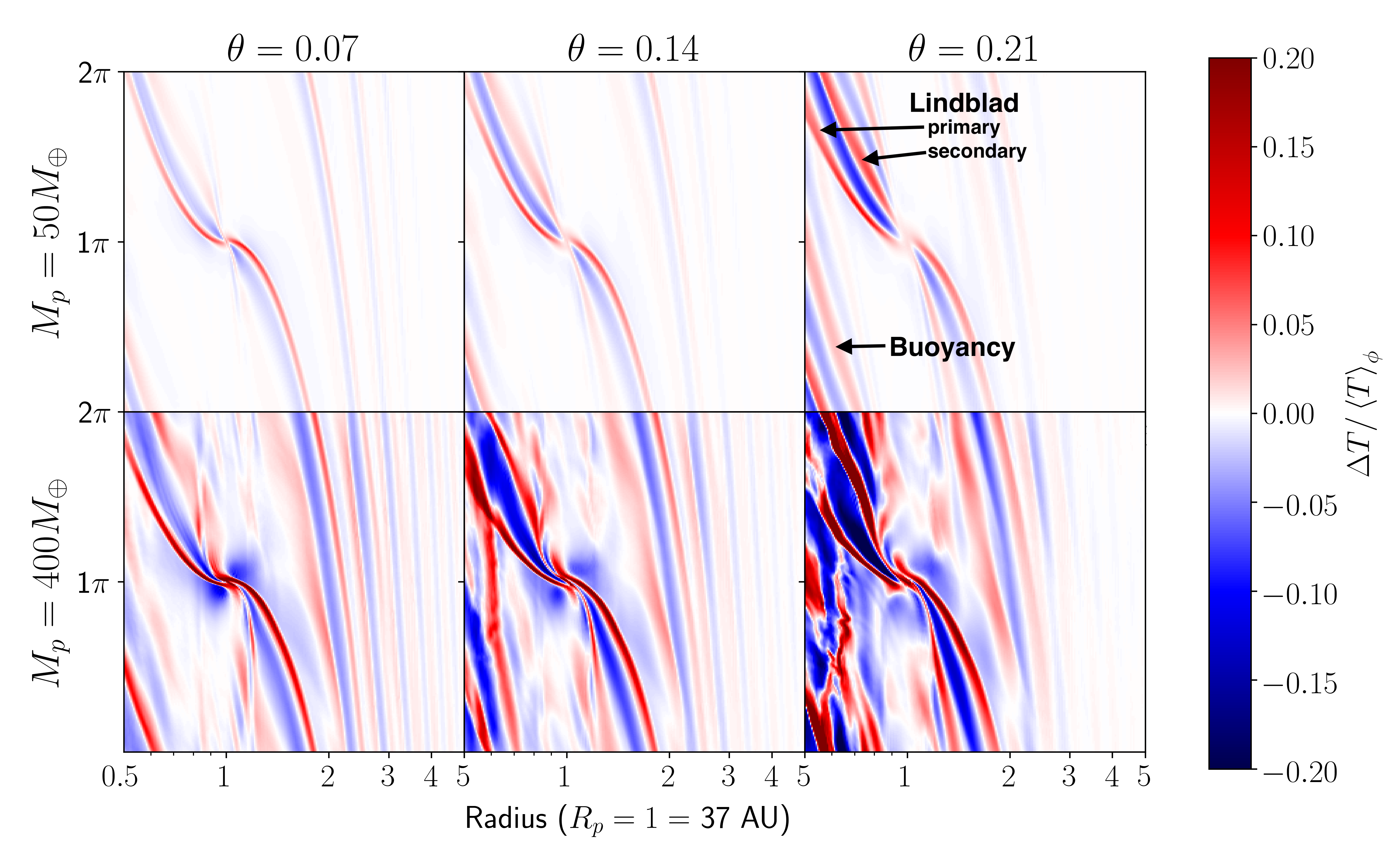}

    \caption{Overall spiral temperature structure for $t_c = 1$ at various polar cuts, with an $x$-axis range of $\{0.5, 5\}$. Secondary Lindblad and buoyancy arms increase in strength with increasing altitude above the midplane, while more tightly-wound arms become  tclearly visible. 
    }
    \label{fig:opacity}
\end{figure*}
\begin{figure*}
    \centering
    \includegraphics[width=0.98\textwidth]{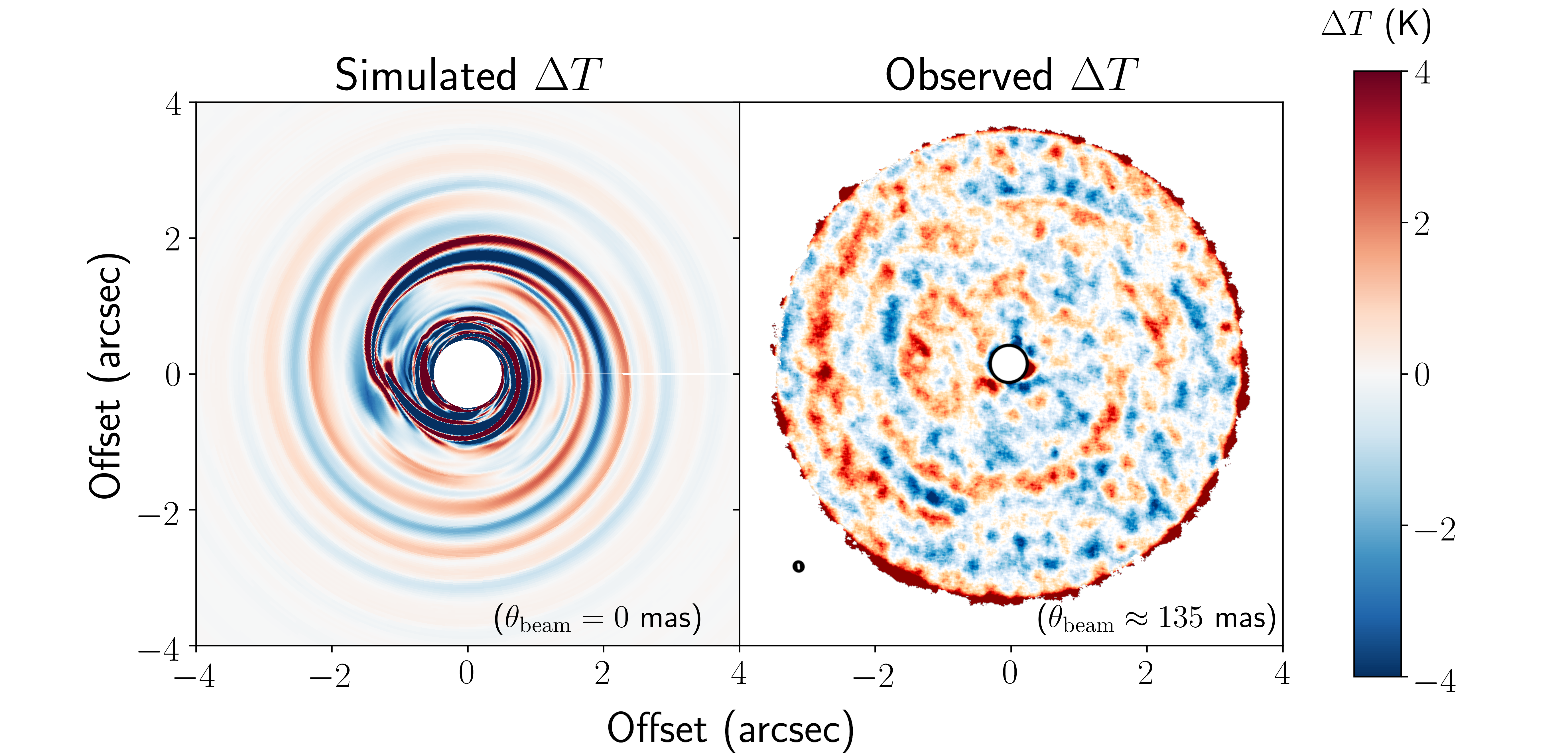}
    \caption{At \textit{left}, our gas temperature structure for $M_p = 400 M_\oplus$ and $t_c = 1$, cut at $\theta = 0.21$ ($\sim3H$) and including only hydrodynamics $PdV$ contributions. We rescale in physical units to compare to the $^{12}$CO brightness-temperature map from \cite{Teague2019}, plotted at \textit{right} (note the beam at the bottom-left of the panel).
    }
    \label{fig:teague}
\end{figure*}

\subsection{Observational diagnostics} \label{sec:observation}
In Figure \ref{fig:opacity}, we plot temperature cuts in the disk at $\theta=1, 2,$ and $3H$; this is intended to qualitatively mimic the different layers of the disk probed by different gas tracers. As before, we normalize and subtract away the average in each azimuthal ring; however, we extend the radial range of the plots to $\{0.5, 5\}$ to emphasize 
that $(\Delta T/\left<T\right>_{\phi})_{\rm spiral}$ is only $\gtrsim 0.025$ within a factor of 2-3 orbital radii from the perturber. We present results from the $t_c = 1$, 50 $M_\oplus$ and 400 $M_\oplus$ models as representative cases. In Figure \ref{fig:teague} we compare our simulated results (scaled to physical units) with those of \cite{Teague2019}.

For both planet masses, the primary Lindblad spiral widens, and a secondary spiral emerges \citep{Bae2018a,Bae2018b}, with increasing altitude in the disk. Buoyancy spirals likewise grow in strength and number with increasing altitude. Both Lindblad spirals wind more tightly with increasing distance from the planet, as expected from our Equation \ref{eq:re_k}---which in the limit $r \gg r_p$, reduces to $\textup{Re}(k_m) \approx (m/\gamma h_{\rm eff} R)(\Omega_p/\Omega)$; convolved with a beam, these two arms emanating from a single point may resemble observations of the CQ Tau \citep{Wolfer2020}. On the other hand, buoyancy spirals are tightly wound even close to the planet. We find, in line with previous isothermal simulations \citep[e.g.,][]{Fung2014,Fung2016}, that planet-carved gaps are circular in the midplane, and widen with increasing altitude; but in a departure from previous work, at higher altitudes our non-isothermal gaps exhibit a substantial ``break'' at the planet location.

\begin{figure}
    \centering
    \includegraphics[width=0.5\textwidth]{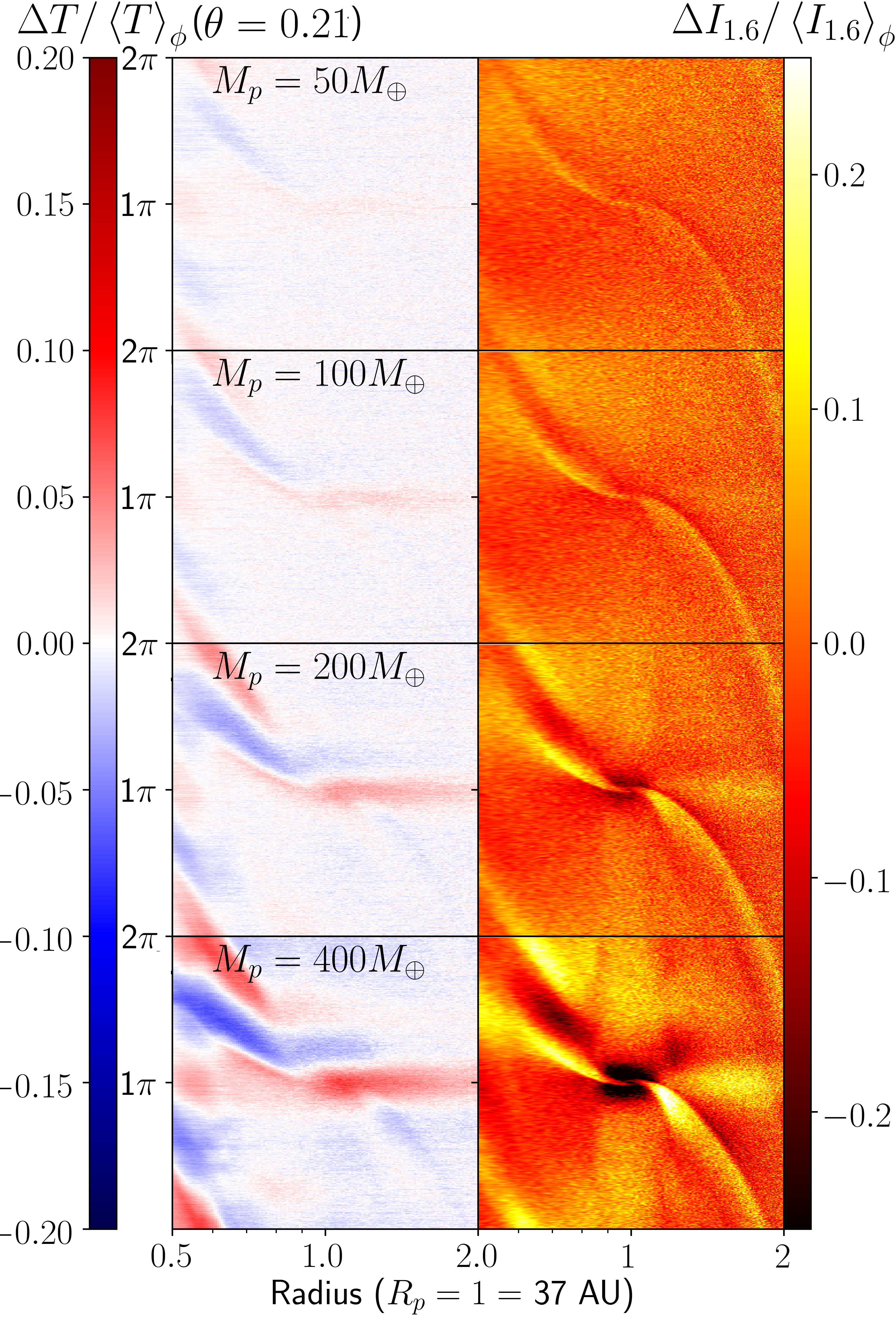}
    \caption{Azimuthal perturbation of temperature cut at $\theta = 0.21$ \textit{(left)} and
    $H$-band ($\lambda \approx 1.6 \mu m$) scattered-light image at face-on in $r-\phi$ coordinates \textit{(right)} from \texttt{HOCHUNK3D} radiative-transfer post-processing (omitting $P dV$ contributions). Arms only become prominent in scattered light for planets with $M_{\rm p}/M_{\rm th} \gtrsim 1$, which substantially alter the scattering surface. 
    }
    \label{fig:radtemp}
\end{figure}

For a better observational understanding, we post-process our hydrodynamical simulations in radiative transfer with \texttt{HOCHUNK3D}, using $10^9$ photons to obtain background temperatures and scattered-light images at face-on for each of our $t_c = 1$ runs. In these simulations, $PdV$ contributions from the hydrodynamics are obviated, and only explicit irradiation effects on $\Delta T$ are included. We plot both in Figure \ref{fig:radtemp}.

For our temperature panels, we plot the $\theta = 0.21 \approx 3H$ layer studied by gas observations, as in Figure \ref{fig:opacity}; taking a cut in this fashion allows us to isolate the effects of spiral shadowing from those of vertical redistribution of material (see \ref{sec:lindblad}, paragraph 2 for a more detailed discussion). In the inner disk, we find azimuthal asymmetries that trace the primary and secondary Lindblad arms, ranging from $1\%$ for our 50 $M_\oplus$ case to $8\%$ for the 400 $M_\oplus$ case whose arms strongly perturb the disk surface. The outer spirals cause no shadowing effect on temperature, but for super-thermal planets, the reduction of disk scale-height near the planet location exposes the outer disk at $\phi = \pi$ to increased stellar irradiation, giving the impression of a radial ``arm'' in temperature. We note that this effect is expected to be transient and disappear once a gap is opened.

While spiral shadowing can be noticeable at high altitudes, it has a markedly lower impact on the disk as a whole. In mass-weighted vertical average, the azimuthal temperature perturbation obtained with \texttt{HOCHUNK3D} is typically $<1-3\%$ interior to the planet, but $<1\%$ in the outer disk. As these differences are consistent with those plotted in Figure \ref{fig:aggregate_lindblad_amplitudes} and \ref{fig:quantitative_p50_perturbation2} for our $t_c = 10^{-2}$ runs, we surmise that they arise primarily from vertical redistribution of disk material in the pre-existing stratified temperature structure, already accounted for in our hydrodynamics. In any case, $P dV$ work in Lindblad spirals (and, at high altitudes and large planet masses, buoyancy spirals as well) overwhelms any radiative effects.

In our near-infrared scattered-light images, the inner Lindblad spiral becomes prominent for the super-thermal $M_p = 200, 400 M_\oplus$ runs; the outer arms are less clearly observable. This aligns with expectations that Lindblad spirals ought to be visible only when planets substantially alter the disk scattering surface, and with the simulations of \cite{Dong2016}, who test substantially higher thermal masses (albeit in isothermal disks with temperature constant along cylinders). In scattered light, the inner primary and secondary Lindblad arms are nearly as strong as each other, but the outer secondary Lindblad and buoyancy arms are barely visible. As in the temperature, there is a radial pseudo-``arm'' in scattered light for super-thermal planets that reshape disk structure around them.

\section{Conclusions}
Using 3D hydrodynamic simulations with an adiabatic equation of state and beta-cooling as well as radiative transfer pre-and post-processing, we quantitatively investigate the spiral temperature perturbations driven by planets in their natal disks, and provide an empirical fit (Equation~\ref{eq:empirical_fit}). Our simulations employ a stratified background temperature, as appropriate for a passively heated disk in hydrostatic equilibrium. In the short-$t_c$ (dimensionless cooling timescale) limit, equivalent to isothermal equation of state, we observe azimuthal temperature perturbations of $<0.5\%$ for exterior Lindblad spirals once vertically averaged (0.3-3\% for interior; depending on planet mass). This effect is mainly caused by a vertical redistribution of material in a stratified background. 

Typical protoplanetary disks are optically thick. We find, with radiative-transfer post-processing) that a secondary irradiation effect---shadowing from the spirals themselves, on top of the stratified background---has only a minor impact on vertically-averaged disk temperature. However, in the tenuous upper layers of the disk probed by $^{12}$CO mapping ($\rho(\theta = 0.21 \approx 3H)/\rho_{\rm mid} \approx 0.01$), shadowing has an effect on \textit{inner} spirals---from 1\% at 50 $M_\oplus$ to a potentially observable 8\% at 400 $M_\oplus$, assuming $t_c = 1$---but remains unimportant for outer spirals.

For longer cooling times, the radiative effects remain, but are overwhelmed by local $P dV$ heating, which produces a vertically-averaged temperature perturbation ranging from 3-11\% for $t_{\rm c} = 1$ and 4-14\% for $t_c = 100$, as mass increases from 50-400 $M_\oplus$. Lindblad spirals are strongest in the midplane, whereas buoyancy spirals are strongest in the low-density disk atmosphere, and so are somewhat weaker in vertical average. Our super-thermal, long-cooling models show azimuthal temperature perturbations comparable to those observed in $^{12}$CO in TW Hydrae \citep{Teague2019} and CQ Tauri \citep{Wolfer2020}, and provide a starting point for simulations with a more realistic, radius-dependent cooling time \citep[e.g.,][]{Miranda2020}.

Our investigations reveal that gas temperature and observations tracing material distributions complement each other as signatures of planet-driven spirals. Hydrodynamical scale-height perturbations are stronger for inner Lindblad spirals \citep[e.g.,][]{Fung2015, Dong2017} and for short cooling times \citep[][]{Ziampras2020} enhancing their visibility in scattered light. By contrast, outer Lindblad spirals are geometrically larger, and spirals with longer cooling times experience $P dV$ heating, making them more prominent in temperature. Spiral temperature structure typically persists for 2-3 windings inside and outside the planet's orbital radius before fading to an azimuthal contrast of ${\sim}2.5\%$, even for super-thermal mass planets
; this, conversely, allows us to use observed arms to constrain the location of the perturber. 

Buoyancy waves, an inherently 3D phenomenon requiring a non-isothermal equation of state, have historically been found to be weak \citep{Zhu2012,Lubow2014,Zhu2015} in simulations where temperature is constant along cylinders. We find that a realistic, stratified temperature structure amplifies the effect of buoyancy (see also \cite{Bae2021}, whose greater vertical temperature gradient leads to even stronger buoyancy resonances), which fundamentally is a process that causes hot material to rise and expand and cold material to sink and contract. Relatively unimportant near the midplane, buoyancy spirals strengthen in the hot disk atmosphere, becoming comparable in $\Delta T/T$ to the more extensively studied Lindblad spirals despite a weaker density perturbation. 

\acknowledgments
We thank Jaehan Bae, Kees Dullemond, Logan Francis, Cassandra Hall, Nienke van der Marel, and Zhaohuan Zhu for useful discussions. We thank the referee for a helpful report that improved the quality of this manuscript. Numerical computations were performed on infrastructure provided by WestGrid and Compute Canada. RD is supported by the Natural Sciences and Engineering Research Council of Canada and the Alfred P. Sloan Foundation. JF gratefully acknowledges support from the Institute for Advanced Study.

\appendix
\section{Temperature fit}
\label{sec:appendix}

We fit the radially and vertically stratified temperature profile from \texttt{HOCHUNK3D} for a passively heated disk in hydrostatic equilibrium (see section \ref{sec:diskmodel}) as the following:
\begin{equation} \label{eq:t_rtheta}
    T_{\rm fit}(r, \theta) = (T_{\rm atm}(r) - T_{\rm mid}(r))(1 - f_2(r, \theta) - f_4(r, \theta)) + T_{\rm mid}(r)
\end{equation}
where

\begin{equation}\label{eq:t_mid}
    T_{\rm mid}(r) = T_{\rm mid, 0}\left(\frac{r}{R_p}\right)^{\beta_{\rm mid}}
\end{equation}
is the midplane temperature,
\begin{equation}
    T_{\rm atm}(r) = T_{\rm atm, 0}\left(\frac{r}{R_p}\right)^{\beta_{\rm atm}}
\end{equation}
is the temperature in the optically thin upper part of the disk atmosphere, and 
\begin{equation}
    f_2 = (1 - f_{4, 0}(r)) \exp \left(-\theta^2/\theta_2^2(r)\right)
\end{equation}
\begin{equation}
    f_4 = f_{4, 0}(r) \exp \left(-\theta^4/\theta_4^4(r)\right)
\end{equation}
are interpolations between the two. We obtain that $T_{\rm mid, 0} = 31.522 {\rm K}$, $\beta_{\rm mid} = \-0.384$, $T_{\rm atm, 0} = 106.655 {\rm K}$, and $\beta_{\rm atm} = -0.422$. We fit $f_{4, 0}$, $\theta_2$, and $\theta_4$ as second-order polynomials in the logarithm of radial position $r$, 

$$g(r) = a_0 \left(\frac{r}{R_p}\right)^{a_1 + a_2 \log_{10} (r/R_p)} $$. 

\begin{figure}[h]
    \centering
    \includegraphics[width=0.94\textwidth]{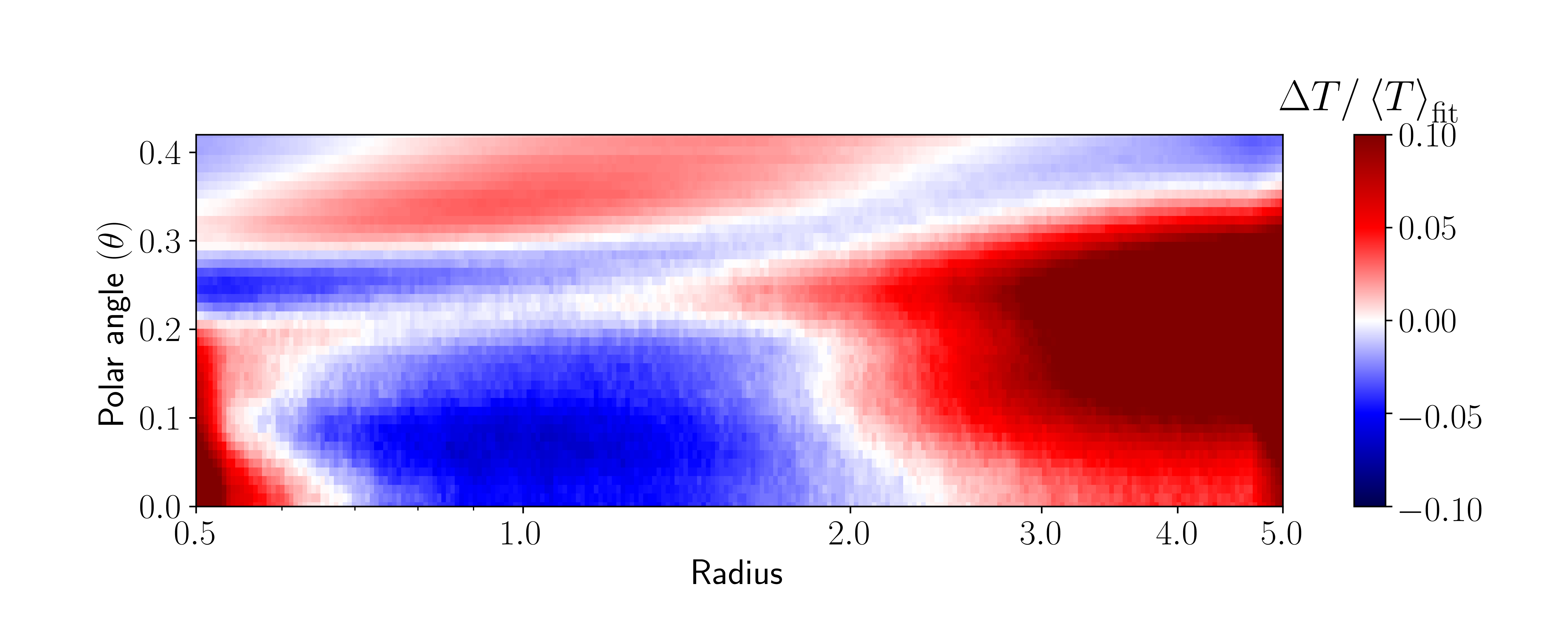}
    \caption{Temperature error at each radial-polar location between the RT simulation and our parametrized fit. Typically, this is within 5\% in our range of interest ($0.5 < r < 2$), showing that our model can capture well the temperature structure of realistic disks. We emphasize that both the RT output and fitted model are both axisymmetric, so the deviation plotted here is not with respect to some azimuthal average.
    }
    \label{fig:temp_error}
\end{figure}

For $\theta_2$, $a_0 = 0.13127$, $a_1 = -0.28745$, and $a_2 = 1.63323$; for $\theta_4$ we have $a_0 = 0.30463$, $a_1 = 0.16974$, $a_2 = -0.06711$; and for $f_{4, 0}$, $a_0 = 0.91957$, $a_1 = -0.079279$, $a_2 = 0.20098$. In Figure \ref{fig:temp_error}, we plot the relative error $\Delta T/T_{\rm fit}(r, \theta)$, where in this case 
\begin{equation}
    \Delta T/T_{\rm fit}(r, \theta) = \frac{1}{2\pi} \int_0^{2\pi} d\phi \left(\frac{T_{\rm actual}(r, \theta, \phi)}{T_{\rm fit}(r, \theta)} - 1\right)
\end{equation}
\newpage
\section{Temperature and density plots}
\label{sec:temp_dens_plots}
\begin{figure*}[h]
    \centering
    \includegraphics[width=0.98\textwidth]{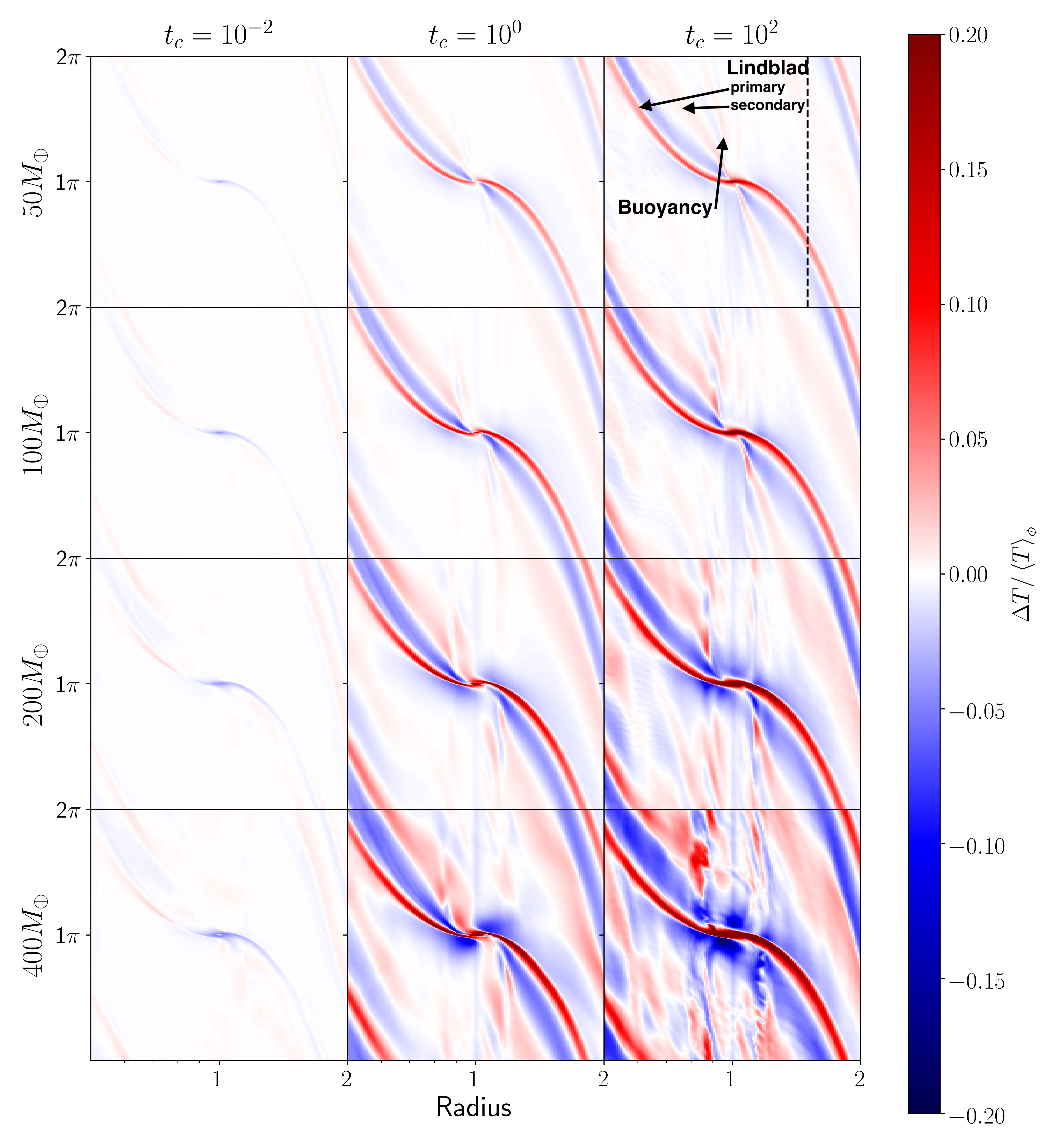}
    \caption{Azimuthal perturbation in vertically-averaged temperature obtained from \texttt{PEnGUIN} hydrodynamics simulations of our model grid. Perturbation amplitude, for both the Lindblad and buoyancy spirals, increase with planetary mass and cooling times. The vertical line at $r= 1.5$ in our $M_p = 50 M_\oplus$, $t_c = 10^2$ run indicates the cut we take for our temperature plot in the upper panel of Figure \ref{fig:quantitative_p50_perturbation}. All runs were taken at 25 orbits.
    }
    \label{fig:vert_temp_struct}
\end{figure*}
\newpage
\begin{figure*}
    \centering
    \includegraphics[width=0.98\textwidth]{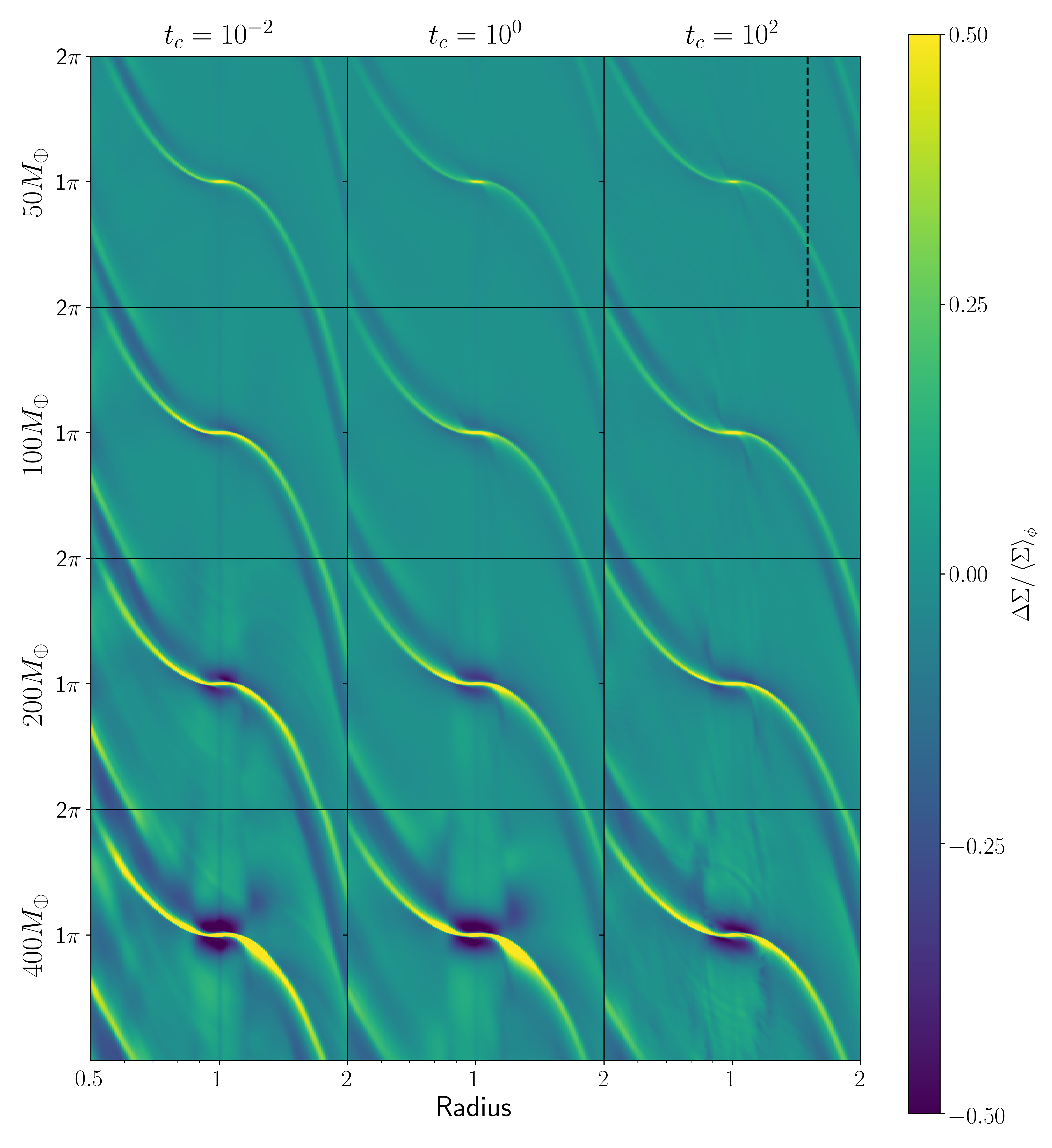}
    \caption{Azimuthal perturbation of surface density in our simulations. These are clearly strongest in the isothermal limit, weakening somewhat with increasing cooling time at a given planet mass. The straight line at the planet radius counterbalances concentration of material at the planet location, which is strongest for high planet masses and short cooling times \citep[as in e.g.,][]{Fung2019}. The vertical line at $r = 1.5$ in our $M_p = 50 M_\oplus$, $t_c = 10^2$ run indicates the cut we take for our temperature plot in the upper panel of Figure \ref{fig:quantitative_p50_perturbation}.
    }
    \label{fig:sigma_plot}
\end{figure*}
\newpage
\begin{figure*}
    \centering
    \includegraphics[width=0.98\textwidth]{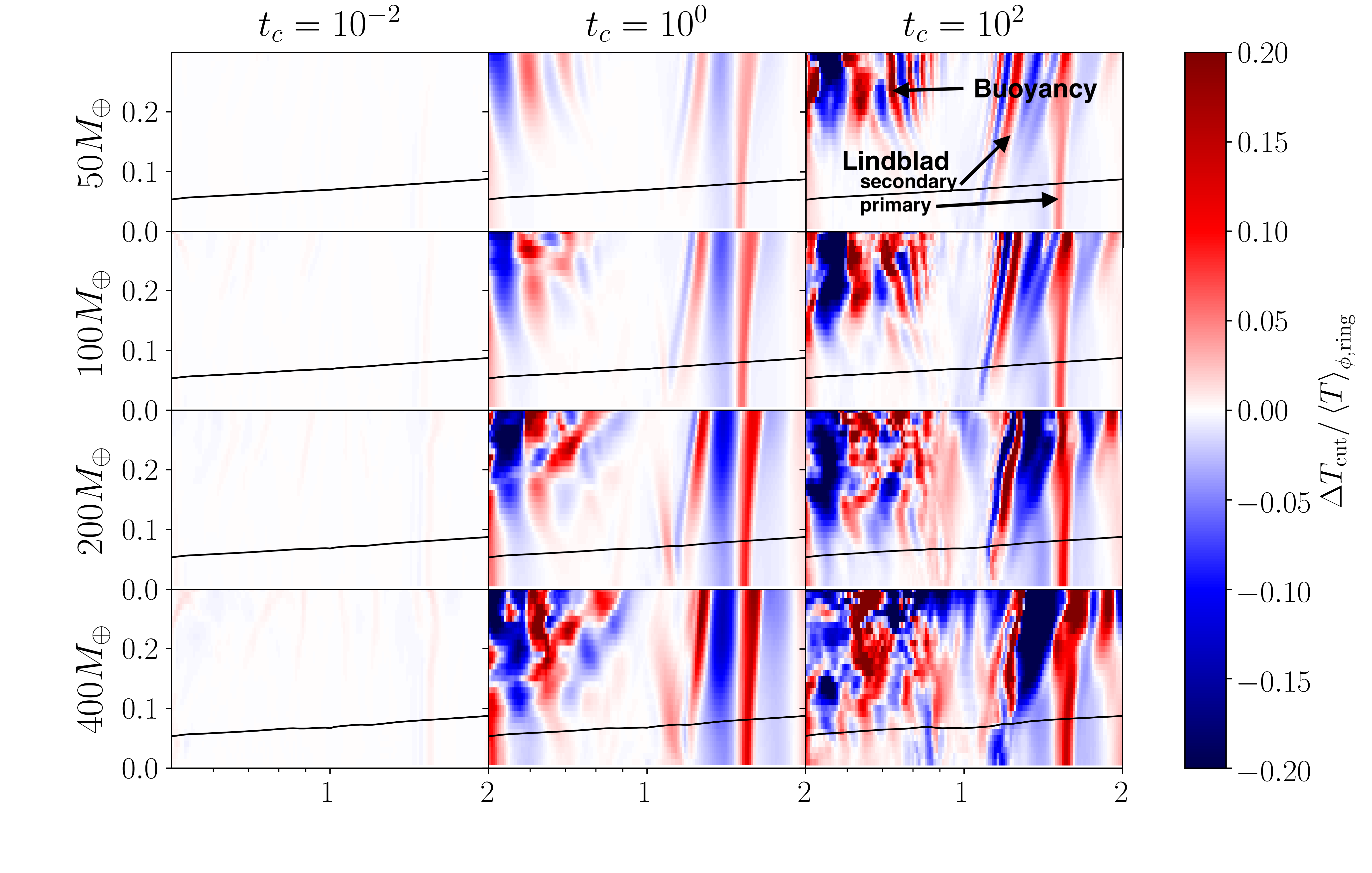}
    \includegraphics[width=0.98\textwidth]{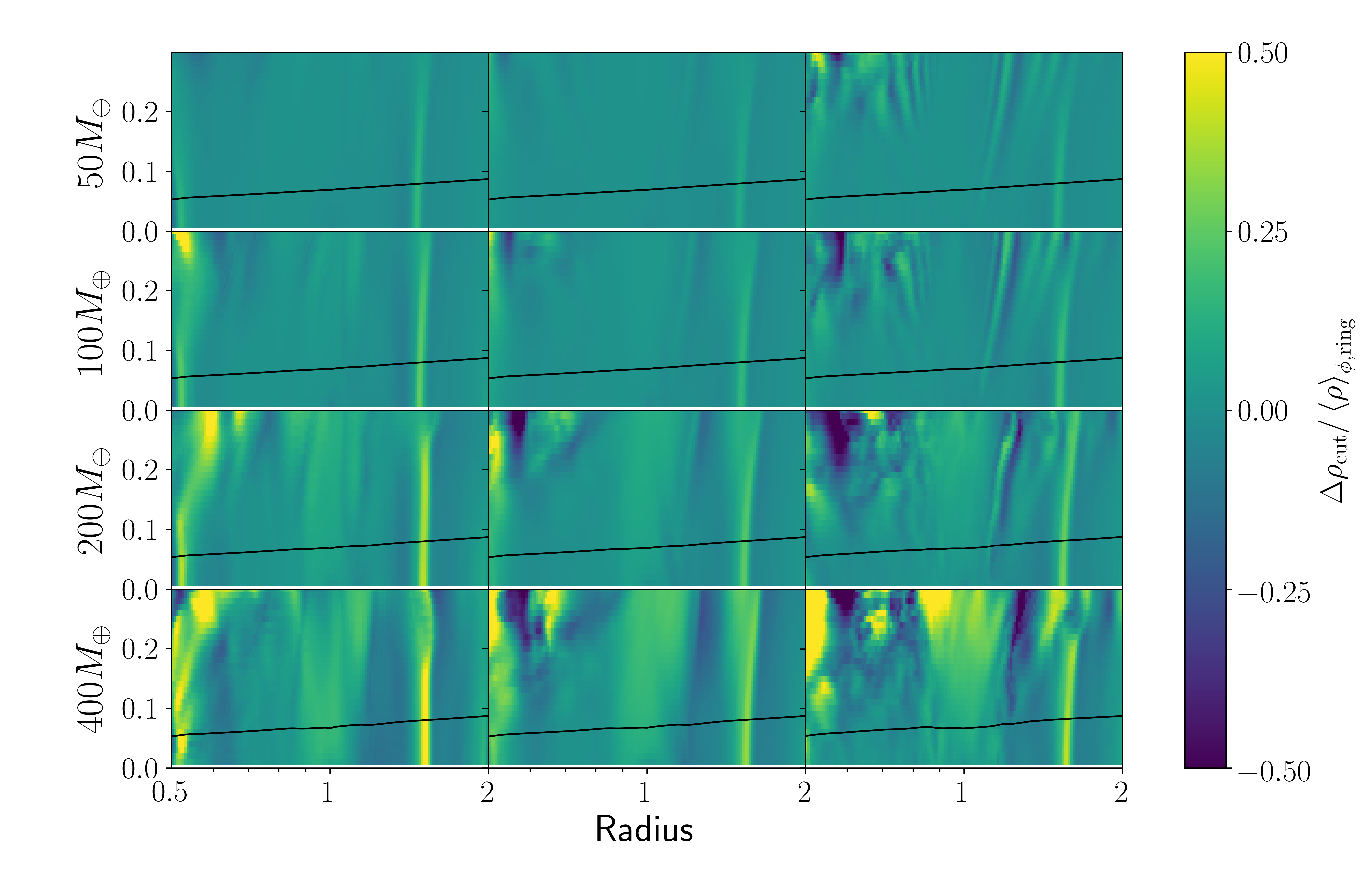}
    \caption{Plot of the temperature perturbation $\Delta T_{\rm cut}$ \textit{(above)} and density perturbation $\Delta \rho_{\rm cut}$ \textit{(below)} in a cut at $\phi = -\pi/2$ relative to the planet, relative to the azimuthal averages $\left<T\right>_{\phi, \rm ring}$ and $\left<\rho\right>_{\phi, \rm ring}$ at each ($r,\theta$). Spherical radius $r$ is on the $x$-axis and polar angle $\theta$ on $y$. Lindblad spirals bend outward because the vertical increase of temperature shifts the effective resonance location outward \citep{Artymowicz1993}. Buoyancy resonances are also distorted, especially for super-thermal planets which themselves meaningfully alter the disk temperature structure.
    }
    \label{fig:temp_dens_aziavg_diff}
\end{figure*}

\newpage
\newpage
%\bibliography{sample63}{}

%\bibliographystyle{aasjournal}



\end{document}